\documentclass{icrc}

\usepackage{times}
\usepackage{graphicx} 

\begin{document}

\title{Mean Square Radius of EAS Electrons}
\author[1]{R.\,I. Raikin}
\author[1]{A.\,A. Lagutin}
\affil[1]{Department of Theoretical Physics, Altai State University, Barnaul, Russia}
\author[2]{N. Inoue}
\affil[2]{Department of Physics, Saitama University, Saitama, Japan}
\author[3]{A. Misaki}
\affil[3]{Advanced Research Institute for Science and Engineering, Waseda University, Tokyo, Japan}

\correspondence{\\R.\,I.~Raikin (raikin@theory.dcn-asu.ru)}

\runninghead{R.\,I.~Raikin et al.: Mean Square Radius of EAS Electrons}
\firstpage{1}
\pubyear{2001}

\maketitle
\sloppy

\begin{abstract}
Detailed theoretical study of the mean square radius of extensive air
shower electrons has been made in connection with further development of
scaling formalism for electron lateral distribution function. 
A very simple approximation formula, which allows joint description
of all our results obtained in wide primary energy range and for
different observation depths is presented. The sensitivity of the mean
square radius to variations of basic parameters of hadronic interaction
model is discussed.
\end{abstract}

\section{Introduction}
In a series of publications [\citet{Raikin3,Raikin1,Raikinsc}, see also
\citet{contr2}] we described model-independent scaling property of lateral
distribution function (LDF) of electrons in the air showers. These results
allow to reproduce the electron LDF up to $r\sim (25 - 30)R_{\rm m.s.}$,
where $R_{\rm m.s.}$ is mean square radius of electrons. In particular, it
was shown that if one use the mean square radius as a radial scale
parameter of lateral distribution function instead of Moliere unit, then
normalized electron LDF becomes invariable on the primary energy and the
age of cascade. In other words, the dependence of the shape of lateral
distribution function on energy, observation depth, chemical composition of
primaries and even features of hadronic interactions is well described by
the variation of single scale parameter $R_{\rm m.s.}$. Unfortunately, such
a sensitive characteristic can not be estimated well directly from
experimental data so far as it demands high precision measurements of
lateral distribution of electron component in wide radial distance range
(from several meters to several thousands meters from the core). Thus the
detailed theoretical study of $R_{\rm m.s.}$, including its sensitivity to
the hadronic interaction model, is necessary for the comparison of our LDF
with experimental data of different arrays.

In this paper we present our latest results on the mean square radius of
electrons in extensive air showers (EAS). Detailed analysis of these
results shows that there is a possibility of very simple joint description
of all data obtained for primary energies $E_0=(10^{14}-10^{18})$~eV and
atmospheric depths $t=(614-1030)~{\rm g/cm}^2$. We present such formula
here together with initial results on the sensitivity of $R_{\rm m.s.}$ to
variations of basic parameters of hadronic interaction model.

\section{Calculation methods}
\hspace*{\parindent}In order to simulate extensive air showers of superhigh
energies we resort to the combination of full Monte-Carlo treatment of
hadronic cascade with analytical expressions of the results of numerical
calculations of electromagnetic subshowers. Two different approaches were
used to obtain the characteristics of pure electromagnetic cascade
initiated by primary photon: the semi-analytical Monte-Carlo method and the
method based on the numerical solution of adjoint cascade equations. For
detailed description of these methods see
\citet{plya1,scn2,adjoint2,Raikin1}. It should be noted here that the main
advantage of semi-analytical Monte-Carlo method is high accuracy in
tracking very distant particles, because complete Monte-Carlo treatment is
applied for low-energetic part of cascade with threshold energy $E_{\rm
th}=0.1$~MeV. On the other hand, the numerical solution of adjoint cascade
equations taking into account the deflection of photons in the multiple
Compton scattering is preferable for independent calculations of moments of
lateral distribution function (such as mean square radius) and gives very
high performance in fluctuation problem and sensitivity problem.

For hadronic part of the cascade we used simplified algorithmic generator
\citep{Raikin3,Raikin1} based on the quark-gluon string model
\citep{Kaid2,Shab2} with extrapolation to $10^{18}$~eV. Though our
generator is very simple, especially in comparison with modern
high-developed codes, the main features of interactions, that make a major
impact on shower development (including fluctuations of full and partial
inelasticity coefficients) are well reproduced by our program [see
\citet{Raikin1,RaikinTH}]. Besides this code is well adapted for the
investigation of sensitivity of results to variations of basic parameters
of hadron-nucleus interaction model.

Due to complete Monte-Carlo treatment of hadronic cascade, the computation
time depends strongly on primary energy. Nevertheless, the simplicity of
hadronic generator together with rejection of tracking muons in combination
with analytical expressions used for electromagnetic component allowed us
to perform very speeding calculations, that ensure maximum 1.5\% relative
statistical error in $R_{\rm m.s.}$ estimation.

Thus we calculated cascade curves and lateral distributions of electrons
in $(10^{14}-10^{18})$~eV proton-induced EAS at six atmospheric depths
from 614 to 1030~$\rm g/cm^2$ in standard atmosphere. Since we implemented
analytical expressions of nor\-malized average LDF and mean square radius
for each partial electromagnetic subshower taking into account fluctuations
of shower size caused by both hadronic and electromagnetic parts of a
cascade, the mean square radius of EAS was calculated in the following way:
\begin{equation}
R_{\rm m.s.}=\left[\frac{\sum\limits_{i}\left(N_e R^2_{\rm m.s.}\right)_i}{\sum\limits_{i}
\left(N_e\right)_i}\right]^{1/2},\label{rms_calc}
\end{equation}
were index~$i$ denotes characteristics obtained for $i^{\rm th}$ individual
extensive air shower, the sum goes over all simulated EAS of fixed primary
energy.

\section{Results}

\begin{figure}[t]
\vspace*{2.0mm} 
\includegraphics[width=8.3cm]{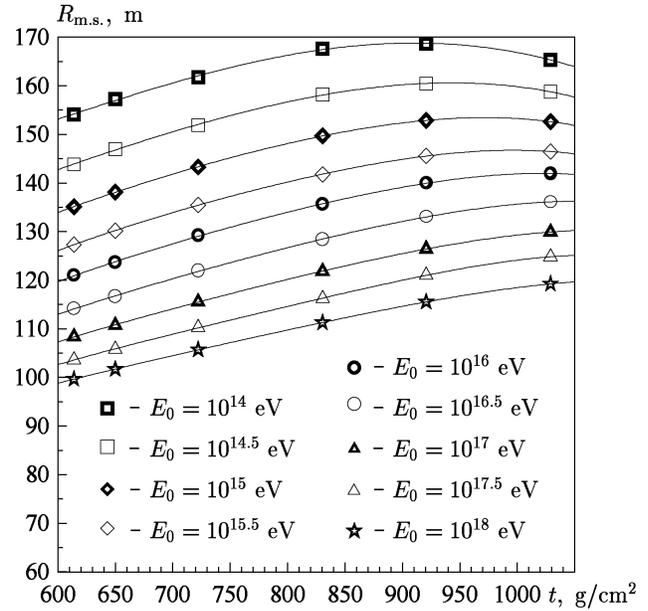} 
\caption{Mean square radius of electrons in vertical proton-initiated EAS.
Symbols~-- our calculational data for $E_0=(10^{14}-10^{18})$~eV
and $t=(614-1030)~\rm g/cm^2$. Curves~-- polynomial approximation}
\label{rms(t)}
\end{figure}

\hspace*{\parindent}The results of our calculations of the mean square radius
of electrons in vertical extensive air showers generated by primary protons
of $E_0=(10^{14}-10^{18})$~eV for atmospheric depths
$t=(614-1030)~{\rm g/cm}^2$ are presented in Fig.~\ref{rms(t)}.
One can see from this figure, that at fixed atmospheric depth $R_{\rm
m.s.}$ decreases with primary energy. That is, our results indicate
narrowing lateral distribution of electrons with primary energy in whole
considered energy range. Though {\em narrowing rate} $\rm NR=\partial
R_{\rm m.s.}/\partial \log E_0$ becomes smaller when energy increases, it
is still essential at $10^{18}$~eV. It can be understood as a reaction to
the decrease of ages of powerful partial electromagnetic subshowers, which
give main contribution to the EAS electron yield.

The behaviour of mean square radius with atmospheric depth is, however,
more complicated. The basic tendency is obvious expansion of shower in
process of development realized in increasing mean square radius.
Nevertheless, there is distinct maximum in $R_{\rm m.s.}(t)$ obtained for
relatively low primary energies, that shifts deeper into the atmosphere
when energy increases. The analysis of our results shows that such a
behaviour can not be explained by only the influence of the increase of air
density. There is another one mechanism of the narrowing of average lateral
distribution related to the correlation between shower size [$(N_e)_i$] and
width [$(R_{\rm m.s.})_i$] in individual showers at fixed primary energy.

\begin{figure}[t]
\vspace*{2.0mm} 
\centering
\includegraphics[width=8.3cm]{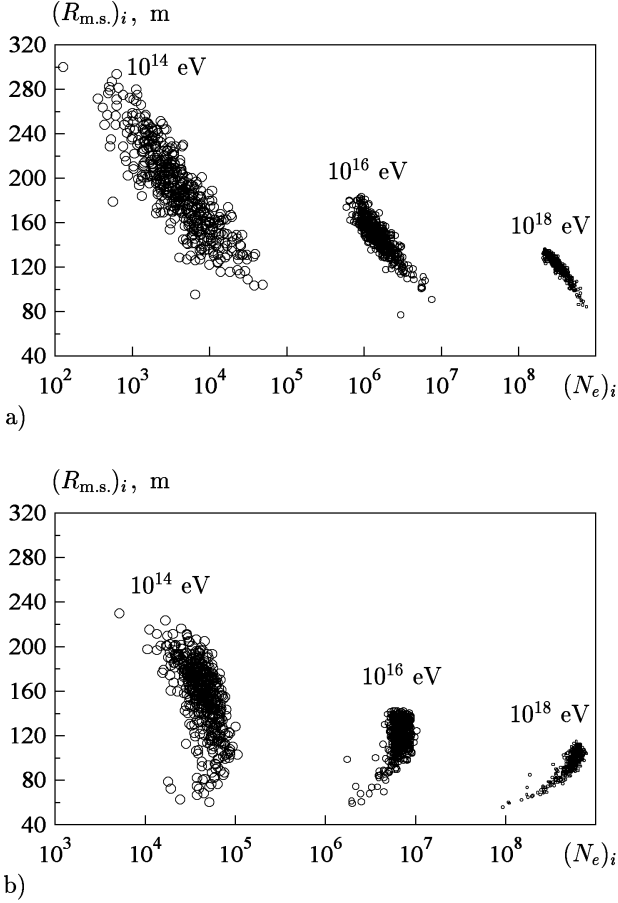} 
\caption{Correlation between $(N_e)_i$ and $(R_{\rm m.s.})_i$ in
individual extensive air showers
of $E_0=10^{14},~10^{16}$ and $10^{18}$~eV at $t=1030~\rm g/cm^2$~(a) and
$t=614~\rm g/cm^2$~(b)}
\label{Rms(Ne)corr}
\end{figure}

In Fig.~\ref{Rms(Ne)corr} we show this correlation by plotting $(R_{\rm
m.s.})_i$ vs. $(N_e)_i$ at sea level~(a) and at 614~$\rm g/cm^2$~(b) for
individual showers of $10^{14}$,~$10^{16}$ and $10^{18}$~eV (500 showers of
each energy were included in data set). It is seen from part a) of a figure
that at sea level $(R_{\rm m.s.})_i$ and $(N_e)_i$ demonstrate strong
anticorrelation. Since relatively large and simultaneously narrow
individual showers of considered energy give more essential contribution to
average LDF [see also eq.~(\ref{rms_calc})], the expansion of the average
shower is slowing down with increasing fluctuations when atmospheric depth
increases. As analysis shows, such anticorrelation exists at observation
depths located deeper than level $t_{\rm opt}$, where fluctuations of
$(N_e)_i$ are minimal (according to our results, $t_{\rm opt}=t_{\rm
max}+55~\rm g/cm^2$, where $t_{\rm max}$ is depth of maximum of average
cascade curve). Thus we should expect existence of such a maximum at higher
energies in case of artificial showers, which continue their development
below the sea level assuming the further exponential growth of air density
together with increasing fluctuations. On the other hand, for atmospheric
depths above $t_{\rm opt}$ such as, for example, 614~$\rm g/cm^2$ for
$E_0=10^{18}$~eV (see Fig.~\ref{Rms(Ne)corr}b) mean square radius correlate
with shower size and this mechanism works conversely.

In this connection it is necessary to note, that $R_{\rm m.s.}$ estimated
as~(\ref{rms_calc}) describes the shape of average lateral distribution of
electrons and deviates distinctly from {\em average mean square
radius} $\left< R_{\rm m.s.}\right>$, which one can calculate using data
about individual showers~as
\[
\left< R_{\rm m.s.}\right>=\displaystyle\frac{1}{n}\sum_{i=1}^n (R_{\rm
m.s.})_i.
\]
The difference becomes essential when observation level is located far from
$t_{\rm opt}$. For example, at sea level $10^{14}$~eV individual
showers are (on average) significantly wider than average shower
($\left< R_{\rm m.s.}\right>$ is 15\% larger than $R_{\rm m.s.}$).

\begin{figure}[t]
\vspace*{2.0mm} 
\includegraphics[width=8.3cm]{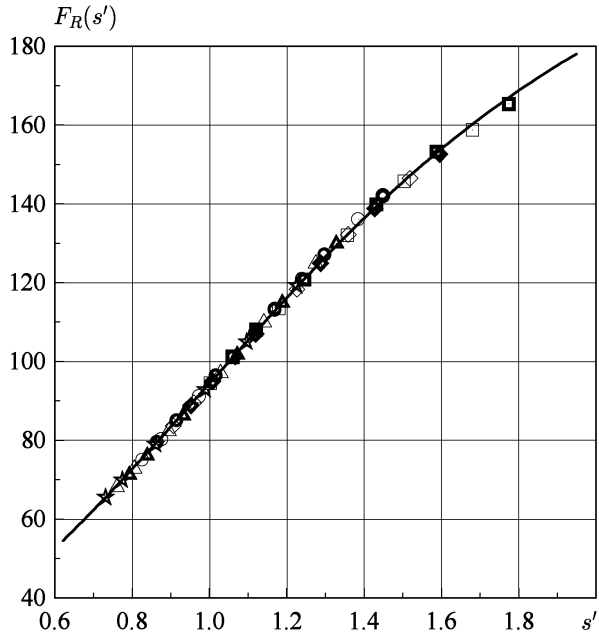} 
\caption{Function $F_R(s')=R_{\rm m.s.}(s')\times
\rho_0/\rho(t)$, where $s'=t/(t_{\rm max}+100~{\rm g/cm^2})$.
Symbols are same as in Fig.~\ref{rms(t)}. Curve~-- fitting
function~(\ref{Rms_app})}
\label{Rms_sc}
\end{figure}

It is important that, though mean square radius 
is affected by number of different factors, a very simple approximation
of all our results based only on the information about longitudinal
shower development is possible. On Fig.~\ref{Rms_sc} the whole set
of our calculational data is presented in the form
$F_R(s')=R_{\rm m.s.}(s')\times
\rho_0/\rho(t)$, where $\rho(t)$ is air density at depth $t$,
$\rho_0=1.225~\rm g/cm^3$, $s'=t/(t_{\rm max}+100~{\rm g/cm^2})$.
It is seen that functions $F_R(s')$ for different primary energies
are precisely superposed with each other in overlapping $s'$ intervals.
The fit of $F_R(s')$ with maximum
relative error of~1\% was obtained using method of maximum likelihood as
$F_R(s')=a+b\cdot{\rm arctg}(s'-1)$, where $a$ and $b$~-- free parameters.
Thus we can propose an analytical expression for $R_{{\rm
m.s.}}(E_0,t)$ as follows:
\[
R_{{\rm m.s.}}(E_0,t)=\frac{\rho_0}{\rho(t)}\,173.0\times
\]
\begin{equation}
\times\left[0.546
+\frac{2}{\pi}~{\rm arctg}\left(\frac{t}{t_{\rm
max}+100~{\rm g/cm^2}}-1\right)\right],~{\rm m}.\label{Rms_app}
\end{equation}
Our results on $t_{\rm max}$ for proton-induced showers were
approximated~as
\begin{equation}
t_{\rm max}(E_0)=740+65\,\lg (E_0/10^{18}~\rm eV),~g/cm^2.\label{tmax}
\end{equation}

Assuming the validity of the superposition model for showers initiated by a
nucleus of energy $E_0$ and mass $A$, we obtain the relation
\[
R^A_{\rm m.s.}(E_0,t)=R^p_{\rm m.s.}(E_0/A,t),
\]
where upper indexes denote the type of primary particle.

\section{Discussion}

\hspace*{\parindent}Expression~(\ref{Rms_app}) together with scaling
parameterization of LDF \citep{Raikin_Brazil,contr2} allows one to obtain
the reliable data on normalized average lateral distribution function of
EAS electrons up to $r\sim (25-30)R_{\rm m.s.}$ ($\sim 2000-3000$~m) in
wide primary energy interval and for any practically important atmospheric
depth from mountain level to sea level. Nevertheless, there is always vital
question related with applicability of any theoretical result in the field
of EAS research: how strong it depends on hadronic interaction model used
in calculations?

In our papers [see, for example, \citet{Raikin3,Raikin1,RaikinTH}] it was
shown that scaling property of electron LDF is very poorly sensitive to
variation of basic parameters of model. Thus mean square radius carries the
whole information about the influence of features of hadronic interactions
on the shape of LDF as well as about mass composition and shower
development. Since the key parameters of hadronic interactions most
influential for the longitudinal development of EAS are inelastic cross
section
of \mbox{$p$-air} collisions and inelasticity \citep{Knapp}, 
we have performed the basic test of sensitivity of
expression~(\ref{Rms_app}) by the change of these parameters in our code.
We used~\mbox{$p$-air} cross sections from different well-known hadronic
models \citep{Knapp} (including MOCCA'92 cross section, which demonstrates
highest increasing rate with primary energy), and independently varied
inclusive energy spectra of secondaries in such a way that
spectrum-weighted moments of inclusive spectra change by $\pm 20\%$ [see
\citet{Bat2}]. Our results show, that within the limits defined by
widespread hadronic models formula~(\ref{Rms_app}) remains valid.

It is also important, that if one extrapolate expression~(\ref{Rms_app}) to
ultrahigh primary energies ($E_0> 10^{18}$~eV), then values of
variable $s'$ corresponding to $t\approx (920-1030)~\rm g/cm^2$ will be
kept within interval well presented in our data set by the values
calculated for moderate energies and higher observation levels [for
example, $s'=0.95$ when $E_0=10^{20}$~eV and $t=920~\rm g/cm^2$, assuming
the validity of formula~(\ref{tmax})]. Though such extrapolation is rather
formal, we expect that it can give reliable predictions at least in the
energy range, where Landau-Pomeranchuk-Migdal and magnetic field effects of
hadron-initiated showers and also $\pi^0$-air inelastic interactions are
negligible. Of course, special calculations are needed to confirm the
reliability of extrapolation of formula~(\ref{Rms_app}).

\balance

\section{Conclusion}

\hspace*{\parindent}This paper represents the further
development of scaling formalism \citep{Raikin3,Raikin1,Raikinsc} for the
description of lateral distribution function of EAS electrons. The
calculations of the mean square radius of electrons in extensive air
showers have been made for primary energies $E_0=(10^{14}-10^{18})$~eV and
atmospheric depths $t=(614-1030)~{\rm g/cm}^2$. Very simple approximation
formula for $R_{\rm m.s.}(E_0,t)$, that gives the one-valued relation
between depth of maximum of average cascade curve and the shape of average
lateral distribution of electrons is presented. It can be practically
implemented during comparisons of experimental data of large ground-based
air shower arrays with results obtained by fluorescent detectors.

\begin{acknowledgements}
One of the authors (R.\,I.\,R.) is greatly indebted to Japan-Russia Youth
Exchange Center (JREX) for support of his research work in Japan in form of
a fellowship. He also wishes to thank Profs. A.\,Misaki and N.\,Inoue and
all members of Saitama University cosmic ray research group for the warm
hospitality.
\end{acknowledgements}

\end{document}